\documentclass[journal]{IEEEtran}
\ifCLASSINFOpdf
\else
\fi
\usepackage{footnote}
\usepackage{epstopdf}
\usepackage{subfigure}
\usepackage{flushend}
\usepackage{multirow}
\usepackage{setspace}
\usepackage{color,graphicx}
\usepackage[square,comma,sort&compress,numbers]{natbib}
\usepackage{amsmath}
\usepackage{amssymb}
\DeclareMathSizes{10}{9}{7}{6}
\IEEEaftertitletext{\vspace{-2\baselineskip}}
\newtheorem{theorem}{Theorem}
\newtheorem{lemma}{Lemma}
\newtheorem{proposition}{Proposition}

\begin{document}
\title{Raptor Codes in the Low SNR Regime}
\author{{Mahyar~Shirvanimoghaddam,~\IEEEmembership{Member,~IEEE,}
and Sarah~Johnson,~\IEEEmembership{Member,~IEEE}}

\thanks{The material in this paper has been submitted in part to the IEEE International Communications Conference (ICC), Kuala Lumpur, Malaysia, May 2016. http://arxiv.org/abs/1510.07728.

M. Shirvanimoghaddam and S. Johnson are with School of Electrical Engineering and Computer Science, The University of Newcastle, NSW, Australia (e-mail: \{mahyar.shirvanimoghaddam, sarah.johnson\}@newcastle.edu.au).}}

\maketitle
\begin{abstract}
In this paper, we revisit the design of Raptor codes for binary input additive white Gaussian noise (BI-AWGN) channels, where we are interested in very low signal to noise ratios (SNRs). A linear programming degree distribution optimization problem is defined for Raptor codes in the low SNR regime through several approxiamtions. We also provide an exact expression for the polynomial representation of the degree distribution with infinite maximum degree in the low SNR regime, which enables us to calculate the exact value of the fractions of output nodes of small degrees. A more practical degree distribution design is also proposed for Raptor codes in the low SNR regime, where we include the rate efficiency and the decoding complexity in the optimization problem, and an upper bound on the maximum rate efficiency is derived for given design parameters. Simulation results show that the Raptor code with the designed degree distributions can approach rate efficiencies larger than 0.95 in the low SNR regime.
\end{abstract}
\IEEEpeerreviewmaketitle

 \section{Introduction}
\label{Sec-Intro}
\IEEEPARstart{I}{nspired} by Shannon's revolutionary work and his seminal paper ``A Mathematical Theory of Communication'' published in 1948, a fundamental question has been posed in communications: \emph{how can we efficiently and reliably transmit information?} Shannon also gave a basic answer: \textit{coding can do it}. Since then, how to find the practical coding scheme that can approach the fundamental limits established by Shannon has been at the heart of coding theory and communications. This paper provides a practical framework to design a capacity approaching code for an additive white Gaussian noise (AWGN) channel in the low SNR regime.
\subsection{Fountain Codes}
Fountain codes \cite{Fountain} are a class of codes on graphs designed mainly for reliable transmission over an erasure channel with unknown erasure probability, which can generate a potentially limitless number of coded symbols for a given set of information symbols of size $k$. The decoder of a fountain code can then recover the original $k$ information symbols from any set of $m>k$ received coded symbols with high probability. Fountain codes are mainly designed such that $m$ is close to $k$ for any erasure rate and the decoding time is close to linear in $k$ \cite{RaptorBSC}.

The Luby Transform (LT) code \cite{Luby} was the first practical realization of Fountain codes, where the degree of each coded symbol was obtained from a single probability mass function, referred to as the degree distribution. In LT codes, the Soliton distribution or robust Soliton distribution \cite{MRSD} was originally used as the degree distribution; however, a capacity approaching LT code was shown to have an average degree increasing with the message size with a logarithmic slope, which makes it very hard to design a linear time encoder/decoder \cite{RaptorBSC}. A more practical extension of LT codes, namely Raptor codes, were proposed by Shokrollahi in \cite{Raptor}, which is a simple concatenation of a high rate low density parity check code (LDPC) and an LT code. The encoding and decoding of these codes are linear in terms of the  message length; thus practical for several applications with large data transmission.

Fountain codes were also studied for unequal error protection (UEP) in \cite{UEP}, intermediate recovery of message symbols for any given fraction of coded symbols in \cite{InterGrowth}, and minimizing reception overhead for any given instantaneous decoding state in \cite{OnlineFountain}. Due to their excellent performance on erasure channels, Raptor codes have been recently adopted in the 3GPP MBMS standard for broadcast file delivery and streaming services \cite{MBMS3GPP}, the DVB-H IPDC standard for delivering IP services over DVB networks, and DVB-IPTV for delivering commercial TV services over an IP network \cite{DVB-IP}. More generally, Raptor codes were shown to be an excellent candidate for forward error correction in the application layer, where the goal is to recover from symbol erasures (rather than errors) \cite{AL-FEC}.

LT and Raptor codes can also be used over other binary input channels \cite{ratenoisy}. The authors in \cite{RaptorBSC} studied Raptor code design for binary memoryless symmetric channels, where a Gaussian approximation was adopted to derive a simple linear programming optimization problem for the degree distribution design. A lower bound on the fraction of output symbols of degree 2 for a capacity approaching code was also proposed, which was shown to be channel dependent for binary input channels, unlike the erasure channel which allows channel independent capacity approaching codes. Accordingly, the authors in \cite{RaptorBSC} proved that there is no universal Raptor code for binary input AWGN (BI-AWGN) channels and binary symmetric channels (BSC). An adaptive degree distribution for Raptor codes over BI-AWGN channels was designed in \cite{Raptor_WideSNR,global_raptor}, where the transmitter changes the degree distribution after sending a certain numbers of coded symbols. This approach has been shown to achieve rate efficiencies close to 90\% for SNRs larger than -10 dB, where the optimization failed to provide good degree distributions for lower SNRs.  In this paper, we study Raptor codes in the low SNR regime, where we design the degree distribution function and also formulate the asymptotic degree distribution function when SNR goes to zero and the maximum degree goes to infinity.

\subsection{Motivations and Main Contributions}
Many communication systems such as wireless ad-hoc networks, sensor networks, and ultra wide band networks, work under low-SNR conditions per node in the network, although the available degrees of freedom is large \cite{BurtRelay}.  As shown in \cite{FadingLowSNR} in the slow-fading scenario, once a channel is in deep fade, channel coding may not help to increase the reliability of the transmission. If available, cooperative transmission can dramatically improve the performance by creating diversity using the antennas available at the other nodes of the network \cite{WSNLowSNR}. However, many applications do not have access to such helpers in the network to improve reliability. For example, in cellular-based machine-to-machine communications, where a large number of low-power devices randomly transmit their data to a single base station, many devices may be in deep fade due to their constant movement. For this type of network, dedicating resources, such as bandwidth or even relays, to devices is not efficient due to the random bursty nature of the transmission \cite{Mahyar_TWC}. These necessitate the study of efficient transmission strategies for low-power communication systems in the low SNR regime, where increasing the transmission power or using cooperative techniques are not possible and channel coding is the only choice.

Very low rate error control codes are also required for the error reconciliation stage of quantum key distribution systems (QKD). QKD is the first commercial application of quantum physics, which aims to provide a completely secure random key between two disjoint parties. More specifically, in continuous variable (CV) QKD systems, low SNRs are essential for the security of the system. Iterative error control codes have the potential to substantially improve the efficiency of QKD systems. For instance, repeat-accumulate (RA) codes have been investigated to use in CV-QKD systems in \cite{Sarah_J_RA_QKD}, where the codeword length is 64800 and the lowest code rate is 1/20. Although, RA and LDPC codes have been proved in \cite{RAAWGN} to achieve the Shannon limit in AWGN channels when the coding rate approaches zero, in the low-rate region, both RA codes and LDPC codes suffer from performance loss and extremely slow convergence speed with iterative decoding. Multi-edge type LDPC (MET-LDPC) codes were shown in \cite{PhysRevA} to theoretically approach high efficiencies in low SNRs. Parallel-concatenated zigzag-Hadamard (PCZH) codes are also proposed in \cite{ConcZigZagHadamard} which offer the bit-error-rate (BER) of $10^{-5}$ at $E_b/N_0=-1.2$ dB, which is only around 0.4 dB away from the Shannon limit. In \cite{RepeatZigZagHadamard}, a new class of low-rate codes called repeat-zigzag-Hadamard (RZH) codes, were proposed which are essentially serially concatenated turbo-like codes where the outer code is a repetition code and the inner code is a punctured ZH code. Low-rate LDPC codes with simple protograph structure was also proposed in \cite{DivsalarLowRate}, which  can be viewed as hybrid concatenations of simple modules such as accumulators, repetition codes, differentiators, and punctured single parity check codes. Existing very low rate codes, however have a very poor word error rate performance and so will only produce good efficiencies for theoretical threshold results or for finite-length results where the allowed word error rate is very high.

In this paper, we study Raptor codes over BI-AWGN channels in the low SNR regime. We use the general framework proposed in \cite{RaptorBSC} for the degree distribution design of the LT code in the low SNR regime. We show that a simple linear program can be designed for the degree distribution optimization problem in the low SNR regime, which is in fact independent of channel SNR. An exact expression for the degree distribution function in the low SNR regime in the asymptotic case, where the maximum degree goes to infinity, is derived and shown to be a function of the elementary EXIT function, previously defined in \cite{SPA_IT}. We also characterize the optimal fractions of low degrees through approximations of the proposed linear program. Using this approach, we characterize the code efficiency, defined as the ratio of the maximum achievable code rate and  the channel capacity. A simple upper bound is also proposed for the code efficiency. Similarly, the average code degree is also characterized and a simple lower bound is proposed. A more practical degree distribution design is also proposed by considering both the rate efficiency and decoding complexity, where a set of optimal degree distribution functions are derived for different maximum code degrees.

\subsection{Organization}
The rest of the paper is organized as follows. Section II presents an overview of Raptor codes. In Section III, we investigate the Raptor code design in the low SNR regime. Section IV provides  practical degree distribution design for Raptor codes in the low SNR regime and some bounds on the maximum rate efficiency of the code and the decoding complexity. Simulation results are shown in Section VI. Finally Section VII concludes the paper.

\section{Preliminaries}
\label{Sec-Overview}
\subsection{Raptor Code Structure and Definitions}
\begin{figure}[t]
\centering
\includegraphics[scale=0.7]{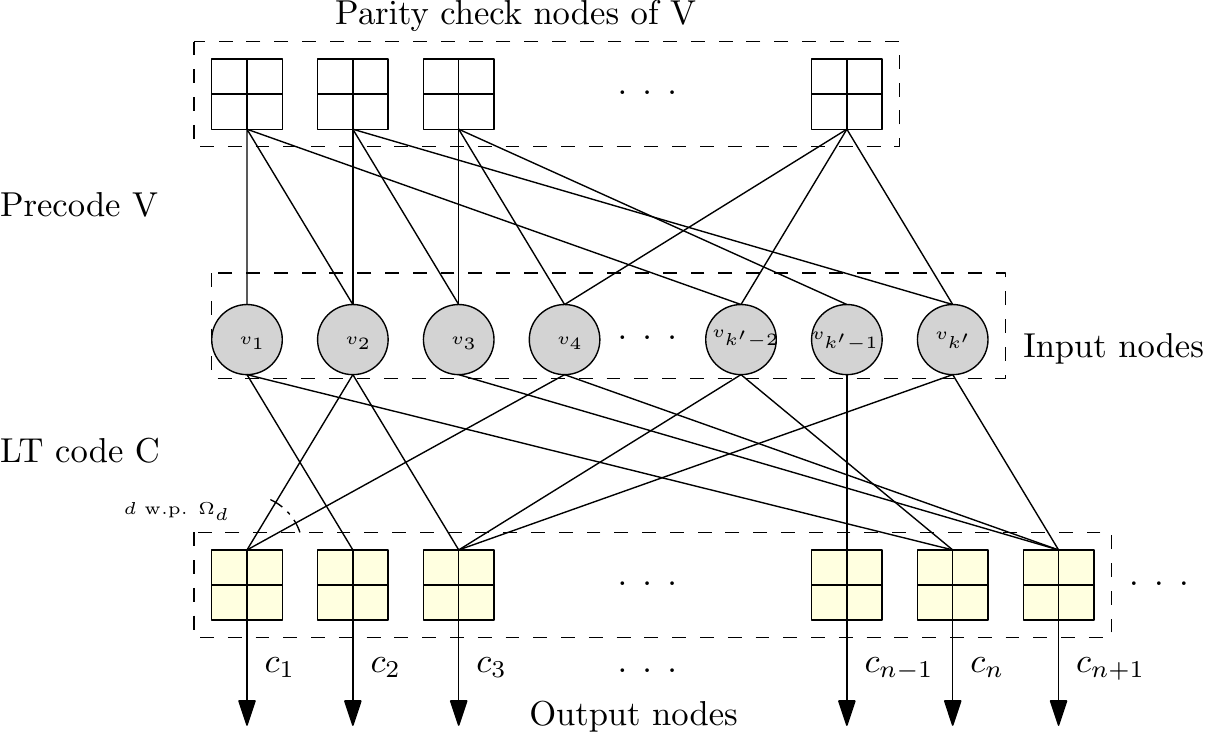}
\caption{Factor graph representation of a Raptor code truncated at codeword length $n+1$.}
\label{LTFIG}
\end{figure}

A binary Raptor code is constructed as the serial concatenation of an outer code $\mathrm{V}$ and an inner LT code $\mathrm{C}$. The outer code is usually a high-rate LDPC code, which encodes a $k$-bit information vector into a $k'$-bit codeword of $\mathrm{V}$, where the codeword bits are referred to as \textit{input symbols}. Then, via an LT code a potentially limitless number of Raptor coded symbols, referred to as \textit{output symbols}, are generated and transmitted over the channel, where each output symbol is the XOR (modulo-2 sum) of a randomly selected fraction of input symbols. More specifically, to generate each output symbol, first an integer $d$, called degree, is obtained from a predefined probability mass function, called the degree distribution function. An output symbol is then generated via the XOR operation on $d$ distinct input symbols, uniformly selected at random. A Raptor code can then be characterized by $(k,\mathrm{V},\Omega(x))$, where $\Omega(x):=\sum_{d}\Omega_dx^d$ denotes the degree distribution polynomial and $\Omega_d$ denote the fraction of output nodes of degree $d$. A Raptor code can be represented by a factor graph, representing both LDPC and LT component codes. Fig. \ref{LTFIG} shows the factor graph of a Raptor code truncated at block length $n$.

The degree distribution can also be defined with respect to edges by $\omega(x):=\sum_{d}\omega_dx^{d-1}$, where $\omega_d$ is the fraction of edges connected to an output symbol of degree $d$. As shown in \cite{Raptor}, $\omega_d$ and $\Omega_d$ can be calculated from each other as follows:
\begin{align}
\omega_d=\frac{d\Omega_d}{\beta},~~~~\Omega_d=\frac{\omega_d/d}{\sum_d\omega_d/d},
\label{outputdd}
\end{align}
where $\beta=\sum_d d\Omega_d$ is the average output node degree. We can also compute the input node and edge degree distributions $\Lambda(x):=\sum_d\Lambda_dx^d$ and $\lambda(x):=\sum_d\lambda_dx^d$, respectively. It has been shown in \cite{RaptorBSC} that in the asymptotic case, when $k$ and $n$ go to infinity, the input node and edge degree distribution follow a Poisson distribution with parameter $\alpha$, i.e.,
\begin{align}
\Lambda(x)=\lambda(x)=\mathrm{e}^{\alpha(x-1)},
\label{inputpoisson}
\end{align}
where $\alpha:=\beta k'/n$ is the average input node degree and $n$ is the number of output symbols. 
\subsection{Decoding of Raptor Codes}
Let $y_i$ denote the channel output for output symbol $c_j$, where $y_j=(-1)^{c_j}+z_j$ and $z_j$ is zero mean Gaussian noise with variance $\sigma^2_z$ for a BI-AWGN channel. The channel signal-to-noise ratio, $\gamma$, is then defined as $\gamma=1/\sigma^2_z$. The channel log-likelihood ratio (LLR) is also defined as follows:
\begin{align}
m^{(0)}_j:=\ln\frac{P(c_j=0|y_j)}{P(c_j=1|y_j)}=\frac{2{}y_j}{\sigma^2_z}.
\label{llroriginal}
\end{align}
It is then easy to show that $m^{(0)}_j$ follows a symmetric Gaussian distribution with mean $2\gamma$, when an all zero codeword is transmitted. Here, a probability density function (pdf) $f(x)$ is called symmetric if $f(x)=\mathrm{e}^xf(-x)$ \cite{Raptor_BIAWGN}.

A sum-product algorithm (SPA) \cite{SPA_IT} can be easily developed for a Raptor code over a binary input memoryless channel on its factor graph representation \cite{RaptorBSC}. In each iteration of the SPA, the LLR messages are passed along each edge between neighboring input symbols and output symbols. Let $m^{(t)}_{o\to i}$ and $m^{(t)}_{i\to o}$ denote messages passing from output node $o$ to input node $i$ and that from input node $i$ to output node $o$, respectively, in the $t^{th}$ iteration of the SPA. The message updating equation at output node $o$ and input node $i$ can be respectively formulated as follows \cite{Raptor_WideSNR}:
\begin{align}
\nonumber m^{(t)}_{o\to i}&=2\tanh^{-1}\left(\tanh\left(\frac{m^{(0)}_o}{2}\right)\prod_{i'\ne i}\tanh\left(\frac{m^{(t-1)}_{i'\to o}}{2}\right)\right),\\
m^{(t)}_{i\to o}&=\sum_{o'\ne o} m^{(t)}_{o'\to i}.
\label{outputupdate}
\end{align}
After a predetermined number of iterations, $T$, the decoded LLR, $m_i$, for input node $i$ is calculated as $m_{i}=\sum_{o} m^{(T)}_{o\to i}$. The decoded LLRs of the input nodes are passed to code $\mathrm{V}$ to recover the original information bits. The entire decoding process is repeated by increasing the number of output symbols, and accordingly their channel outputs, until the $k$-bit information vector is correctly decoded. Note that a decoder verifies the correctness of the decoding by the check equations of the outer code $\mathrm{V}$, and the receiver sends an acknowledgment to the sender via a feedback channel upon the correct verification of the decoding.

We define the realized rate of a Raptor code at SNR $\gamma$, denoted by $R_r(\gamma)$, as the ratio of the message length, $k$, and the average number of output symbols required for the successful decoding of the entire message at the destination. The rate efficiency of the code, denoted by $\eta$, is then defined as the ratio of the realized rate and channel capacity, i.e.,
\begin{align}
\eta(\gamma)=\frac{R_r(\gamma)}{C_{b}(\gamma)},
\label{efficiencyfor}
\end{align}
where $C_b(\gamma)$ is the capacity of a BI-AWGN channel at SNR $\gamma$, which is given by \cite{RaptorBSC}:
\begin{align}
C_{b}(\gamma)=1-\frac{1}{\sqrt{8\pi \gamma}}\int_{-\infty}^{\infty}\log_2\left(1+\mathrm{e}^{-x}\right)\mathrm{e}^{-\frac{(x-2\gamma)^2}{8\gamma}}dx.
\label{biawgncap}
\end{align}
\subsection{Degree Distribution Optimization of Raptor Codes}
In \cite{RaptorBSC}, a framework to construct a Raptor code for a given BI-AWGN channel has been proposed. This framework, referred to as the mean-LLR-exit chart approach, was proposed based on two assumptions. Firstly, the cycle-free assumption which means that the factor graph representation of Raptor codes is locally cycle-free, which is well justified when the graph is large and sparsely connected. This assumption enables us to assume that all incoming SPA messages arriving at a given node are statistically independent. The second assumption is that the probability density of a message passed from an input symbol to an output node along a randomly chosen edge in the graph is a mixture of symmetric Gaussian distributions. This can be satisfied when the degree of each input node is relatively large, then according to the Central Limit Theorem the sum of incoming messages to an input node follows a Gaussian distribution. The mixture of Gaussian model is then due to the irregular degrees of input nodes \cite{Raptor_BIAWGN}.

Since we assume that the message from input node to output node follows a symmetric Gaussian distribution, its pdf can be characterized by its mean. Let $\mu^{(t)}$ denote the mean of message $m^{(t)}_{i\to o}$ along a randomly chosen edge in the $t^{th}$ iteration of SPA. Similar to \cite{Raptor_BIAWGN}, we define a function $f_d(\mu)$, representing the mean of the message from output node to input node, as follows:
\begin{align}
f_d(\mu)=2\mathbb{E}\left[\tanh^{-1}\left(\tanh\left(\frac{Z}{2}\right)\prod_{q=1}^{d-1}\tanh\left(\frac{X_q}{2}\right)\right)\right],
\label{ElemExit}
\end{align}
where $\mathbb{E}$ is the expectation operator, $X_q$, which correspond to the LLR message from input node to output node, is the symmetric Gaussian random variable with mean $\mu$ and $Z$ is a symmetric Gaussian random variable with mean $2\text{SNR}$ representing the channel output LLR. As shown in \cite{RaptorBSC}, under the above assumptions and that the all-zero codeword is transmitted, we have:
\begin{align}
\textstyle\mu^{(t+1)}=\alpha\sum_d\omega_d f_d(\mu^{(t)}).
\end{align}
The goal of Raptor code design is then to find the output node or edge degree distribution to maximize the design rate of the LT code, defined as $R_{\mathrm{design}}:=\beta/\alpha$, such that $\mu^{(t+1)}>\mu^{(t)}$ for $t\ge0$, so that the bit error rate (BER) of input symbols decreases through SPA iterations. For this aim, \cite{RaptorBSC} proposed the following linear program to minimize  $\alpha\sum_{d=1}^{D}\omega_d/d$, which is equivalent to maximizing $R_{\mathrm{design}}$, for given $\alpha$, $D$, and the target maximal message mean $\mu_o$:
 \begin{align}
 \label{optorig}
\text{minimize}&~~\textstyle\alpha\sum_{d=1}^{D}\omega_d/d\\
\nonumber \text{s.t.}~& (i). ~\textstyle\alpha\sum_{d=1}^D\omega_{d}f_d\left(\mu_j\right)>\mu_j, ~~\forall j=1,\cdots,N,\\
\nonumber~&(ii).~\textstyle\sum_{d=1}^{D}\omega_d=1,\\
\nonumber~&(iii).~\textstyle\omega_d\ge 0, ~~\forall d=1,\cdots, D,
 \end{align}
where $\{\mu_j|j=1,\cdots,N\}$ is a set of equally spaced values in range $(0,\mu_o]$ and $\mu_N=\mu_o$. It is important to note that $\mu_o$ is chosen to be large enough to ensure that the decoding of outer code $\mathrm{V}$ is successful.

\section{Raptor Codes in the Low SNR Regime}
In this section, we study Raptor codes in the low SNR regime and an optimal degree distribution function is formulated in the asymptotic case, where the channel SNR goes to zero and the maximum degree goes to infinity.
\subsection{Degree Distribution Optimization in the Low SNR Regime}
Authors in \cite{Raptor_BIAWGN} provided a detailed study of the feasibility of the linear program (\ref{optorig}) based on parameters $\mu_o$ and $\alpha$. More specifically, they showed that for SNRs higher than $\text{SNR}_{\mathrm{low}}=\mu_o/2\alpha$ the linear program is feasible and thus has a solution. The following lemma shows that the linear program (\ref{optorig}) is always feasible in the asymptotic case, where the maximum degree goes to infinity for all SNRs.
\begin{lemma}
\label{feasibility}
The linear program (\ref{optorig}) is feasible for every SNR, when $D$, and accordingly the average input node degree, go to infinity.
\end{lemma}
The proof of this lemma is provided in Appendix \ref{feasibilityproof}.  In the rest of the paper, we assume that the linear program in the low SNR regime is always feasible for an adequately large maximum degree and, due to the linearity of the objective function and all constraints in (\ref{optorig}), the optimization is convex, thus there is always an optimal degree distribution function.

Let us have a closer look at function $f_d(\mu)$ defined in (\ref{ElemExit}), when the channel SNR, $\gamma$, goes to zero. We define the function $h_d(Z,\textbf{X}_d)$ as follows:
\begin{align}
\nonumber h_d(Z,\textbf{X}_d)=2\tanh^{-1}\left(\tanh\left(\frac{Z}{2}\right)\prod_{q=1}^{d-1}\tanh\left(\frac{X_q}{2}\right)\right),
\end{align}
where $\textbf{X}_d\triangleq(X_1,...,X_{d-1})$. Then, it is clear that $f_d(\mu)=\mathbb{E}[h(Z,\textbf{X}_d)]$. We expand this function in a Taylor series about the mean value of $Z$, i.e, $2\gamma$, that is:
\begin{align}
\nonumber &h_d(Z,\textbf{X}_d)\\
\nonumber &=h_d(2\gamma,\textbf{X}_d)+\left(Z-2\gamma\right)\frac{\partial h_d}{\partial Z} +\frac{\left(Z-2\gamma\right)^2}{2}\frac{\partial^2 h_d}{\partial Z^2}+\cdots,
\end{align}
where the derivatives are evaluated at $2\gamma$. By truncating the above equation at the second order terms and taking the expectation with respect to $Z$ from both sides, we have:
\begin{align}
\label{approx1}
\mathbb{E}_Z\left[h_d(Z,\textbf{X}_d)\right]\approx h_d(2\gamma,\textbf{X}_d)+\frac{1}{2}\frac{\partial^2 h_d}{\partial Z^2}\sigma^2_Z,
\end{align}
where $\mathbb{E}[Z]=2\gamma$ and $\sigma^2_Z=4\gamma$, as $Z$ follows a symmetric Gaussian distribution with mean $2\gamma$. The second derivative of $h_d(Z,\textbf{X}_d)$ with respect to $Z$ at $Z=2\gamma$ is given by:
\begin{align}
\left.\frac{\partial^2 h_d}{\partial Z^2}\right|_{Z=2\gamma}=\frac{P_d(1-P_d^2)\tanh(\gamma)(1-\tanh^2(\gamma))}{1-P_d^2\tanh^2(\gamma)},
\end{align}
where $P_d\triangleq\prod_{q=1}^{d-1}\tanh(X_q/2)$. Therefore, (\ref{approx1}) can be rewritten as follows:
\begin{align}
\nonumber \mathbb{E}_Z&\left[h_d(Z,\textbf{X}_d)\right]\approx 2\tanh^{-1}\left(\tanh(\gamma)P_d\right)\\
&+2\gamma\frac{P_d(1-P_d^2)\tanh(\gamma)(1-\tanh^2(\gamma))}{1-P_d^2\tanh^2(\gamma)}.
\label{1stapprox}
 \end{align}
Since $\gamma$ goes to zero, (\ref{1stapprox}) can be further expanded in a Taylor series about zero as follows:
\begin{align}
\nonumber \mathbb{E}_Z\left[h_d(Z,\textbf{X}_d)\right]&\approx 2\gamma P_d +\mathcal{O}(\gamma^2),
\end{align}
which follows from $h_d(0,\textbf{X}_d)=0$, $\left.\frac{\partial^2 h_d}{\partial Z^2}\right|_{Z=0}=0$, and $\left.\frac{\partial h_d}{\partial Z}\right|_{Z=0}=P_d$. Using only the linear terms in the above approximation, we can approximate $f_d(\mu)$ as follows:
\begin{align}
\nonumber f_d(\mu)&=\mathbb{E}_{\textbf{X}}\left[\mathbb{E}_{Z}\left[h_d(Z,\textbf{X})\right]\right]\approx\mathbb{E}_{\textbf{X}}\left[2\gamma\prod_{q=1}^{d-1}\tanh\left(\frac{X_q}{2}\right)\right]\\
&=2\gamma\prod_{q=1}^{d-1}\mathbb{E}\left[\tanh\left(\frac{X_q}{2}\right)\right],
\end{align}
which follows from the fact that $Z$, $X_q$ are mutually independent random variables for $q=1, \cdots, d$. As $X_q$ is assumed to follow a symmetric Gaussian distribution of mean $\mu$, we can define function $\varphi(\mu)$ for $\mu\ge0$ as follows:
\begin{align}
\nonumber \varphi(\mu)&=\mathbb{E}\left[\tanh\left(\frac{X}{2}\right)\right]\\
&=\frac{1}{\sqrt{4\pi \mu}}\int_{-\infty}^{\infty}\tanh\left(\frac{u}{2}\right)\text{e}^{-\frac{(u-\mu)^2}{4\mu}}du.
\label{phifunc}
\end{align}
Thus, $f_d(\mu)$ can be further simplified as follows:
\begin{align}
f_d(\mu)\approx2\gamma\left(\varphi(\mu)\right)^{d-1}.
\end{align}
The left hand side of condition (i) in linear program (\ref{optorig}) can then be rewritten as follows:
\begin{align}
\alpha\sum_{d}\omega_{d}f_d\left(\mu\right)\approx2\gamma\alpha\sum_{d}\omega_{d}\left[\varphi(\mu)\right]^{d-1}.
\end{align}
It is important to note that in the low SNR regime, by using the well-know Maclaurin series of the function $\ln(1+\gamma)$, the AWGN channel capacity can be approximated as follows:
\begin{align}
C_b(\gamma)=\frac{1}{2}\log_2(1+\gamma)=\frac{1}{2\ln(2)}\left(\gamma+\mathcal{O}(\gamma^2)\right).
\label{appcap}
\end{align}
Therefore, for a capacity approaching code (i.e., the design rate is assumed to be very close to the channel capacity) and for a very small $\gamma$, we have:
\begin{align}
\beta=\alpha R_{\mathrm{design}}\approx\frac{\alpha}{2\ln(2)}\left(\gamma+\mathcal{O}(\gamma^2)\right).
\end{align}
Thus by using (\ref{outputdd}), we have:
\begin{align}
\nonumber \alpha\sum_{d}\omega_{d}f_d\left(\mu\right)&\approx{4\ln(2)}\beta\sum_{d}\omega_{d}\left[\varphi(\mu)\right]^{d-1}\\
&={4\ln(2)}\sum_dd\Omega_d[\varphi(\mu)]^{d-1}.
\label{finalcondi}
\end{align}
We can then formulate the optimization problem to maximize the design rate, $R_{\mathrm{design}}$, which is equivalent to maximizing the average output node degree $\beta$, for a given maximum degree $D$, when $\gamma$ goes to zero:
 \begin{align}
 \label{optlowsnr1}
&\text{maximize}~~\textstyle\sum_{d=1}^D d\Omega_d\\
\nonumber&\text{s.t.}~ (i). ~\textstyle\sum_{d=1}^Dd\Omega_{d}[\varphi(\mu_j)]^{d-1}>\frac{\mu_j}{4\ln(2)}, ~~\forall j=1,\cdots,N,\\
\nonumber~&(ii).~\textstyle\sum_{d=1}^{D}\Omega_d=1,\\
\nonumber~&(iii).~\textstyle\Omega_d\ge 0, ~~\forall d=1,..., D.
\end{align}
The above optimization is linear for a given $\mu_o$ and $D$. However, we consider a joint optimization problem, where instead of supplying a predetermined $\mu_o$, the above linear program is converted to a general non-linear optimization problem with the same objective and constraints as in (\ref{optlowsnr1}), but optimizing jointly over $(\mu_o,\Omega(x))$. For this aim, an optimal $\mu_o$ can be found by searching over a relatively large range of discretized values of $\mu_o$, where for each $\mu_o$ the optimization is converted to a linear program. Table \ref{OptDegTabSum} shows the maximum $\mu_o$ for different values of $D$.

\begin{table}[t]
\caption{Optimized degree distributions for different values of the maximum degree $D$.}
\scriptsize
\centering
\begin{tabular}{|c|c|c|c|}
 \hline
$D$&$\mu_o$&$\delta_0$&$\sum d\Omega_d$\\
\hline
\hline
50&16.22&0.0068&6.7579\\
\hline
100&18.75&0.0034&7.6878\\
\hline
200&21.31&0.0017&8.6247\\
\hline
300&22.81&0.0011&9.1700\\
\hline
500&24.71&0.0007&9.8659\\
\hline
1000&27.35&0.0003&10.8052\\
 \hline
\end{tabular}
\label{OptDegTabSum}
\end{table}
\normalsize

\subsection{An Exact Expression for the Optimal Degree Distribution Polynomial in the Low SNR regime}
We first summarize some of the properties of the function $\varphi(x)$ in the following proposition, which are used in the remainder of the paper.
\begin{proposition}
\label{prop1}
The function $\varphi(x)$, defined in (\ref{phifunc}), has the following properties:
\begin{itemize}
  \item[a)] Function $\varphi(x)$ is continuous, concave, and monotonically increasing with $x$.
  \item[b)] The inverse function $\varphi^{-1}(x)$ exists in $[0,1)$, where $\varphi(\varphi^{-1}(x))=x$. The inverse function is continuous, convex, and monotonically increasing with $x$.
  \item[c)] $\varphi(0)=\varphi^{-1}(0)=0$.
  \item[d)] $\lim_{x\to \infty} \varphi(x)=1$.
  \item[e)] $\lim_{x\to 1}\varphi^{-1}(x)=\infty$.
  \item[f)] $\lim_{\delta\to0}\int_{0}^{1-\delta}\varphi^{-1}(t)dt=4\ln(2)$.
\end{itemize}
\end{proposition}
The proof of this proposition is provided in Appendix \ref{prop1proof}. Now, we define $x_j\triangleq \varphi(\mu_j)$ for $j=1,\cdots, N$, and substitute it into constraint ($i$) in (\ref{optlowsnr1}), which is then rewritten as follows:
\begin{align}
\sum_{d=1}^{D}d\Omega_dx_j^{d-1}> \frac{1}{4\ln(2)}\varphi^{-1}(x_j),~~\forall j=1,\cdots,N,
\label{Eq23}
\end{align}
where $x_N=\varphi(\mu_o)$. Eq. (\ref{Eq23}) can then be written in a more general form as follows:
\begin{align}
\label{condnewder}
\sum_{d=1}^{D}d\Omega_dx^{d-1}> \frac{1}{4\ln(2)}\varphi^{-1}(x),~~0\le x\le1-\delta_0,
\end{align}
where $\delta_0=1-\varphi(\mu_o)$. Eq. (\ref{condnewder}) can be also rewritten as
\begin{align}
\label{dercond1}
\Omega'(x)>\frac{1}{4\ln(2)}\varphi^{-1}(x),~~0\le x\le1-\delta_0,
\end{align}
where $\Omega'(x)$ is the derivative of $\Omega(x)$ with respect to $x$. The optimization problem is then to find the optimal degree distribution which jointly maximizes the design rate and minimizes $\delta_0$ (which is equivalent to maximizing $\mu_o$) for a given maximum degree $D$ such that constraints ($i$)-($iii$) in (\ref{optlowsnr1}) are satisfied. By taking the integral of both sides of (\ref{dercond1}) and knowing that $\Omega(0)=0$ and $\varphi^{-1}(0)=0$, we have:
 \begin{align}
 \label{equfinalfor}
\Omega(x)> \frac{1}{4\ln(2)}\int_{0}^{x}\varphi^{-1}(t)dt,~~0\le x\le1-\delta_0.
\end{align}
The following lemma shows how the optimal value of $\delta_0$ changes with the maximum degree $D$.
\begin{lemma}
\label{generallemma}
Let $\delta_D$ denote the minimum value of $\delta_0$ such that (\ref{condnewder}) holds for $0\le x\le 1-\delta_0$ for a given maximum degree $D$ and $\Omega^{(D)}(x)$ denote the respective optimal  degree distribution function. Then the following conditions hold:
\begin{itemize}
\item[1.] $\delta_D>0$ for a finite $D$.
\item[2.] $\delta_D$ is decreasing in $D$.
\end{itemize}
\end{lemma}

The proof of this Lemma is provided in Appendix \ref{generallemmaproof}. As can be seen in Table \ref{OptDegTabSum}, $\delta_D$ decreases with $D$ and it is very close to zero, which is in close agreement with the results of this lemma. The following theorem then gives the optimal degree distribution function when $D\to\infty$ (see Appendix \ref{theorem1proof} for the proof).
\begin{theorem}
\label{theorem1}
The optimal degree distribution function of a Raptor code in the low SNR regime, when $D\to\infty$ is bounded by:
 \begin{align}
 \label{finalOptForT}
\left|\Omega^{(\infty)}(x)- \frac{1}{4\ln(2)}\int_{0}^{x}\varphi^{-1}(t)dt\right|\le \epsilon_{\infty}, ~~0\le x<1,
\end{align}
where $\Omega^{(\infty)}(x)\triangleq\lim_{D\to \infty}\Omega^{(D)}(x)$,
\begin{align}
\epsilon_{\infty}=1-\frac{1}{4\ln(2)}\int_{0}^{1-\delta_{\infty}}\varphi^{-1}(t)dt,
\end{align}
and $\delta_{\infty}\triangleq\lim_{D\to \infty}\delta_D$.
\end{theorem}

It is important to note that according to Lemma \ref{feasibility}, there exists a degree distribution function with a potentially infinite maximum degree for every $\delta>0$, thus, Theorem \ref{theorem1} gives the exact expression for the optimal degree distribution function when $D$ goes to infinity. More specifically, as $\delta_D$ is decreasing with $D$ and there exists at least one degree distribution function for each $\delta>0$, which satisfies condition (\ref{firstcond}), we can conclude that $\lim_{D\to\infty}\delta_D=0$. Therefore, by using Theorem \ref{theorem1} and Lemma \ref{generallemma}, we have $\epsilon_{\infty}=0$ and accordingly the asymptotic degree distribution function, $\Omega^{(\infty)}(x)$ is given by:
\begin{align}
\label{AsymDeg}
\Omega^{(\infty)}(x)=\frac{1}{4\ln(2)}\int_{0}^{x}\varphi^{-1}(t)dt,~~x\in[0,1].
\end{align}
The solid black curve in Fig. \ref{OptDegPolFig} shows the asymptotic degree distribution polynomial for $x\in[0,1]$.
\subsection{Fraction of Output Symbols with Small Degrees}
When $x$ approaches zero, the left hand side of (\ref{AsymDeg}) can be realized as the Maclaurin series of $\frac{1}{4\ln(2)}\int_0^x\varphi^{-1}(t)dt$, which is given by:
\begin{align*}
\frac{1}{4\ln(2)}\int_0^x\varphi^{-1}(t)dt=\frac{1}{4\ln(2)}\sum_{i=1}^{\infty}\frac{x^i}{i!}\left[\frac{d^{i-1}}{dt^{i-1}}\varphi^{-1}(t)\right]_{t=0}.
\end{align*}
The derivatives of an inverse function can be found as follows \cite{todorov1981}:
\begin{align}
\label{inverseDer}
\frac{d^i}{dt^i}\varphi^{-1}(t)=\frac{1}{(\varphi'(t))^{2i-1}}\det\left(\textbf{A}^{(i)}\right),
\end{align}
where the elements of the matrix $\textbf{A}{(i)}$ are given by:
\[A^{(i)}_{k,\ell}=\left\{\begin{array}{ll}
\frac{(k-\ell+1)i-k}{1!}\varphi^{(k-\ell+1)}(t),& 1\le\ell\le\min\{k,i-1\}\\
1,&k=1~\&~\ell=i,\\
0,&\text{otherwise.}
\end{array}\right.\]
The optimized degrees can then be found as follows for $j>1$:

\begin{align}
\Omega_j=\frac{\eta}{4\ln(2)j!}\left[\frac{d^{j-1}}{dt^{j-1}}\varphi^{-1}(t)\right]_{t=0}.
\label{optdegfor}
\end{align}
As shown in (\ref{inverseDer}), to find the optimal degree distribution, we need the derivatives of $\varphi(x)$ at $x=0$. They can be found by using the Maclaurin series of $\tanh(x)$ \cite{Mathhandbook}, as follows:
\begin{align}
\nonumber\varphi(x)&=\frac{1}{\sqrt{4\pi x}}\int_{-\infty}^{\infty}\tanh\left(\frac{u}{2}\right)\text{e}^{-\frac{(u-x)^2}{4x}}du\\
\nonumber&=\frac{1}{\sqrt{4\pi x}}\int_{-\infty}^{\infty}\sum_{n=1}^{\infty}\frac{2(2^{2n}-1)B_{2n}}{(2n)!}~u^{2n-1}\text{e}^{-\frac{(u-x)^2}{4x}}du\\
&=\sum_{n=1}^{\infty}\frac{2(2^{2n}-1)B_{2n}}{(2n)!}~M_{2n-1}(x,2x),
\label{initphifor}
\end{align}
where $B_{2n}$ is the $(2n)^{th}$ Bernoulli number \cite{Mathhandbook} and $M_{2n-1}(x,2x)$ is the $(2n-1)^{th}$ moment of a Gaussian distribution with mean $x$ and variance $2x$, which can be found as follows:
\begin{align}
M_{2n-1}(x,2x)=\sum_{j=0}^{n-1}\dbinom{2n-1}{2j}(2j-1)!! (2x)^{j}x^{2n-2j-1}.
\label{highmom}
\end{align}
The double factorial for odd numbers is defined as $(2j-1)!!=(2j-1)(2j-3)\cdots5.3.1$ and can be also represented as follows:
\begin{align}\label{doublefac}
(2j-1)!!=\frac{(2j)!}{2^jj!}.
\end{align}
By substituting (\ref{highmom}) and (\ref{doublefac}) in (\ref{initphifor}), we have
\begin{align}
\nonumber\varphi(x)&=\sum_{n=1}^{\infty}\sum_{j=0}^{n-1}\frac{2^{j+1}(2^{2n}-1)B_{2n}}{(2n)!}\dbinom{2n-1}{2j}(2j-1)!! x^{2n-1-j}\\
\nonumber&=\sum_{n=1}^{\infty}\sum_{j=0}^{n-1}\frac{2^{2n}-1}{n~j!~(2n-2j-1)!} B_{2n} x^{2n-1-j}\\
&=\sum_{\ell=1}^{\infty}\sum_{n=\lceil \frac{\ell+1}{2}\rceil}^{\ell}\frac{2^{2n}-1}{n(2n-\ell-1)!(2\ell-2n+1)!} B_{2n} x^{\ell},
\label{finalPhiAPp}
\end{align}
where the last equality is derived by replacing $\ell$ instead of $2n-j-1$ and changing the bounds of the summations accordingly. By calculating the first few terms of the right hand side of (\ref{finalPhiAPp}), we have:
\begin{align*}
\varphi(x)=\frac{1}{2}x-\frac{1}{4}x^2+\frac{5}{24}x^3-\frac{13}{48}x^4+\frac{227}{480}x^5+\cdots.
\end{align*}
The $i^{th}$ derivatives of $\varphi(x)$ at $x=0$ can then be calculated as follows for $i\ge1$:
\begin{align}\label{deriviphi}
\varphi^{(i)}(0)=\sum_{n=\lceil \frac{i+1}{2}\rceil}^{i}\frac{(2^{2n}-1)~i!}{n(2n-i-1)!(2i-2n+1)!} B_{2n}.
\end{align}
By substituting (\ref{deriviphi}) in (\ref{inverseDer}) and (\ref{optdegfor}), and due to the fact that $\varphi(0)=0$, the first few optimal values of $\Omega_d$ can be found as follows:
\begin{align*}
\Omega_2&=\frac{1}{4\ln(2)}~=0.3607,\\
\Omega_3&=\frac{1}{6\ln(2)}~=0.2404,\\
\Omega_4&=\frac{1}{24\ln(2)}=0.0601,\\
\Omega_5&=\frac{1}{10\ln(2)}=0.1443.
\end{align*}
Higher degrees cannot be found through this approach as the approximation returns negative values. However, a numerical optimization can be performed for a relatively large maximum degree to obtain the complete degree distribution if required. In Appendix \ref{altproof}, an alternative approach is also proposed to find the optimal degree distribution function in the asymptotic case, when the maximum degree goes to infinity. This approach is based on the fact that the left hand side of (\ref{AsymDeg}) can be realized as a pdf, thus its moments can be easily found. The degree distribution function can then be found through a system of linear equations.

\subsection{Optimal Average Output Node Degree}
As the degree distribution function is a polynomial with non-negative coefficients, all its derivatives are non-negative in $[0,1]$. It is then clear that for a Raptor code in the low SNR regime with the optimal degree distribution function of maximum degree $D$, we have:
\begin{align}
\Omega^{(D)'}(x)\le \Omega^{(D)'}(1),
\end{align}
and due to (\ref{finalOptForT}), we have:
\begin{align}
\Omega^{(D)'}(1)\ge \frac{1}{4\ln(2)}\varphi^{-1}(1-\delta_D).
\end{align}
Therefore, for degree distribution functions of maximum degree $D$, which satisfy (\ref{finalOptForT}) for $x\in[0,1-\delta_D]$, the minimum average degree, denoted by $\beta^{(D)}_{\mathrm{min}}$ is given by:
\begin{align}
\beta^{(D)}_{\mathrm{min}}=\frac{1}{4\ln(2)}\varphi^{-1}(1-\delta_D).
\end{align}

\section{A More Practical Degree Distribution Design}
The asymptotic degree distribution formulated in the previous section is optimal in the theoretical sense, but in practice the performances promised require an infinite number of iterations. It is well knows that practical codes require a gap between curves in the EXIT chart to reduce the number of decoding iterations. In this section, we introduce the parameter $\epsilon$ to force this gap between the optimized degree distribution function and the asymptotic degree distribution function. For the decoder to proceed in iterations, $\epsilon$ should be larger than zero and the higher the value of $\epsilon$ the less number of iterations for the decoder to converge. It also has another effect on the decoders, as larger $\epsilon$ provides higher tolerance to random changes in the convergence of the decoder due to noise. We formulate the optimization problem by taking $\epsilon$ and rate efficiency $\eta$ into account, then provide bounds on the average output node degree, the maximal achievable rate efficiency, and the maximum number of decoding iterations, as a function of $\epsilon$ and $\eta$.

More specifically, we formulate the degree distribution optimization problem as follows, when $\gamma$ goes to zero:
 \begin{align}
 \label{optlowsnr}
&\text{maximize}~~\textstyle\sum_{d=1}^{D}d\Omega_d\\
\nonumber&\text{s.t.}~ (i). ~\textstyle\sum_{d=1}^Dd\Omega_{d}x_j^{d-1}\ge\frac{\eta(\varphi^{-1}(x_j)+\epsilon)}{4\ln(2)}, ~~\forall j=1,\cdots,N,\\
\nonumber~&(ii).~\textstyle\sum_{d=1}^{D}\Omega_d=1,\\
\nonumber~&(iii).~\textstyle\Omega_d\ge 0, ~~\forall d=1,..., D.
\end{align}

\begin{figure}[t]
\centering
\includegraphics[scale=0.34]{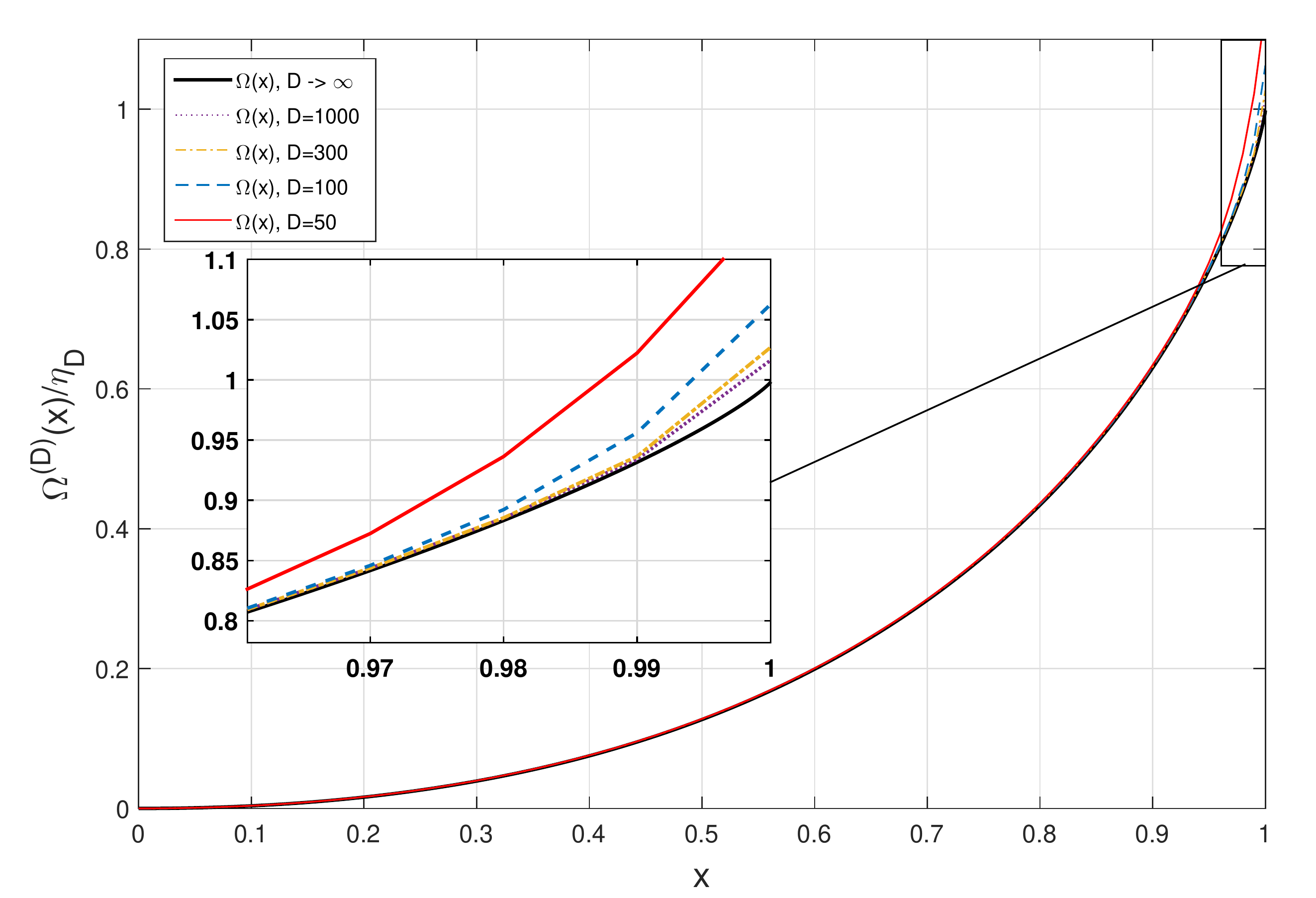}
\caption{Optimal degree distribution functions for low SNR Raptor codes, for different values of D when $\epsilon=0.05$ and $\mu_0=40$.}
\label{OptDegPolFig}
\end{figure}

The above optimization is linear for a given $\eta$, $\epsilon$ and $D$. However, we consider a joint optimization problem, where instead of supplying a predetermined $\eta$, the above linear program is converted to a general non-linear optimization problem with the same objective and constraints as in (\ref{optlowsnr}), but optimizing jointly over $(\eta,\Omega(x))$. For this aim, an optimal $\eta$ can be found by searching over a relatively large range of discretized values of $\eta$, where for each $\eta$ the optimization is converted to a linear program. Note that $\epsilon$ is chosen to be very small, however it can be also set as zero. Table \ref{OptDegTab} shows some optimal degree distributions for different values of $D$ and their corresponding maximum efficiencies, when $\epsilon=0.05$. Fig. \ref{OptDegPolFig} shows the optimal degree distribution polynomials of a Raptor code in the low SNR regime for different values of $D$. As can be seen in this figure, with increasing maximum degree $D$, the degree polynomial gets closer to the optimal degree polynomial found in Theorem 1.

\begin{figure}[t]
\centering
\includegraphics[scale=0.33]{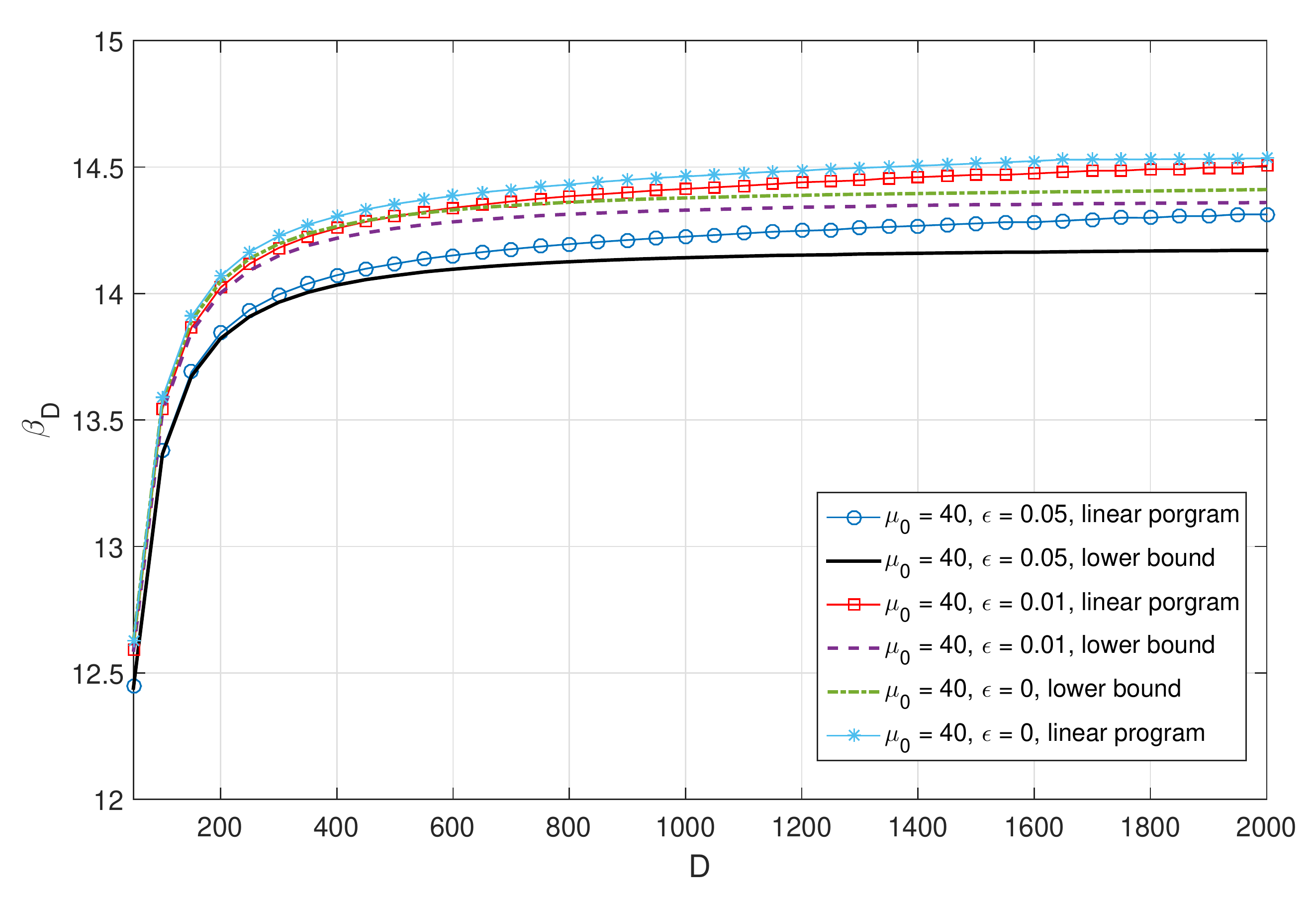}
\caption{The average degree of optimal degree distribution functions versus the maximum degree for different values of $\epsilon$.}
\label{OptBetaFig}
\end{figure}

It is important to note that the above optimization problem will be equivalent to (\ref{optlowsnr1}) for $\epsilon=0$ and $\eta=1$. Also, as can be seen in Table \ref{OptDegTab}, the fraction of low degrees when the maximum degree is large, are very close to those calculated in Section III-C. The following lemma gives a lower bound on the average output node degree as a function of $\mu_o$, $\epsilon$, and $\eta$.
\begin{lemma}
\label{avgdegProp}
The average output node degree for the capacity approaching Raptor code in low SNRs is lower bounded by:
\begin{align}
\sum_{d=1}^{D}d\Omega_d\ge\frac{\eta(\mu_o+\epsilon)}{4\ln(2)}.
\label{avgdegree}
\end{align}
\end{lemma}
The proof of this lemma is provided in Appendix \ref{ProofavgdegProp}. For a sufficiently large value of $\mu_o$, $\varphi(\mu_o)$ is very close to 1, and the lower bound (\ref{avgdegree}) would be very tight. For example, when $\mu_o=40$ and $\epsilon=0.05$, the average output node degree is lower bounded by $14.445\eta$, which is very close to the respective value in Table \ref{OptDegTab}, for different maximum degrees. Fig. \ref{OptBetaFig} shows the average output node degree of Raptor codes with optimal degree distributions versus the maximum degree. As can be seen in this figure, the proposed lower bound is very close to the optimal values, especially in the lower degrees. It is important to note that for a fixed $\mu_o$, the gap between the actual average degree and its lower bound increases with the maximum degree. The reason is that with increasing $D$, condition ($i$) in (\ref{optlowsnr}) can be satisfied for a larger $\mu_o$. One could jointly optimize $\eta$, $\mu_o$, and $\Omega(x)$, so the average degree is this case would be very close to the lower bound with the optimal $\mu_o$ for each $D$.
\begin{table}[t]
\caption{Optimized degree distributions for different $D$ and their corresponding maximum efficiencies, when $\mu_o=40$ and $\epsilon=0.05$.}
\scriptsize
\centering
\begin{tabular}{|c||c|c|c|c|c|c|}
 \hline
$D$&50 &100&200&300&500&1000 \\
\hline
\hline
$\Omega_1$&0.0155 &0.0167&0.0173&0.0174&0.0175&0.0176\\
\hline
$\Omega_2$&0.3140&0.3351&0.3451&0.3488&0.3517&0.3536\\
\hline
$\Omega_3$&0.1764&0.2074&0.2300&0.2309&0.2335&0.2353\\
\hline
$\Omega_4$&0.1508&0.1218&0.0601&0.0695&0.0667&0.0636\\
\hline
$\Omega_5$&-&-&0.1075&0.0873&0.0948&0.1038\\
\hline
$\Omega_6$&-&0.0067&-&0.0002&-&-\\
\hline
$\Omega_7$&-&0.1380&-&0.0805&0.0664&0.0311\\
\hline
$\Omega_8$&0.0599&-&0.0959&0.0004&0.0106&0.0577\\
\hline
$\Omega_9$&0.0970&-&-&-&-&0.0023\\
\hline
$\Omega_{11}$&-&-&-&0.0191&0.0539&-\\
\hline
$\Omega_{12}$&-&-&-&0.0518&0.0087&0.0088\\
\hline
$\Omega_{13}$&-&-&-&-&-&0.0463\\
\hline
$\Omega_{15}$&-&-&0.0341&-&-&-\\
\hline
$\Omega_{16}$&-&-&0.0307&-&-&-\\
\hline
$\Omega_{17}$&-&0.0474&-&-&-&-\\
\hline
$\Omega_{18}$&-&0.0356&-&-&-&-\\
\hline
$\Omega_{20}$&-&-&-&-&0.0427&-\\
\hline
$\Omega_{22}$&-&-&-&-&-&0.0211\\
\hline
$\Omega_{23}$&-&-&-&0.0123&-&0.0133\\
\hline
$\Omega_{24}$&-&-&-&0.0310&-&-\\
\hline
$\Omega_{38}$&-&-&0.0291&-&-&-\\
\hline
$\Omega_{39}$&-&-&0.0078&-&-&-\\
\hline
$\Omega_{41}$&-&-&-&-&0.0251&0.0056\\
\hline
$\Omega_{42}$&-&-&-&-&-&0.0151\\
\hline
$\Omega_{50}$&0.1864&-&-&-&-&-\\
\hline
$\Omega_{59}$&-&-&-&0.0220&-&-\\
\hline
$\Omega_{60}$&-&-&-&0.0020&-&-\\
\hline
$\Omega_{85}$&-&-&-&-&-&0.0021\\
\hline
$\Omega_{86}$&-&-&-&-&-&0.0097\\
\hline
$\Omega_{100}$&-&0.0913&-&-&-&-\\
\hline
$\Omega_{101}$&-&-&-&-&0.0028&-\\
\hline
$\Omega_{102}$&-&-&-&-&0.0109&-\\
\hline
$\Omega_{200}$&-&-&0.0424&-&-&-\\
\hline
$\Omega_{211}$&-&-&-&-&-&0.0026\\
\hline
$\Omega_{212}$&-&-&-&-&-&0.0038\\
\hline
$\Omega_{300}$&-&-&-&0.0268&-&-\\
\hline
$\Omega_{497}$&-&-&-&-&0.0022&-\\
\hline
$\Omega_{498}$&-&-&-&-&0.0125&-\\
\hline
$\Omega_{995}$&-&-&-&-&-&0.0033\\
\hline
$\Omega_{996}$&-&-&-&-&-&0.0031\\
\hline
\hline
$\beta$&12.4457&13.3772&13.8436&13.9957&14.1176&14.2248\\
\hline
\hline
$\eta$&0.8612&0.9253&0.9569&0.9668&0.9741&0.9790\\
\hline
\end{tabular}
\label{OptDegTab}
\end{table}
\normalsize

\begin{figure}[t]
\centering
\includegraphics[scale=0.33]{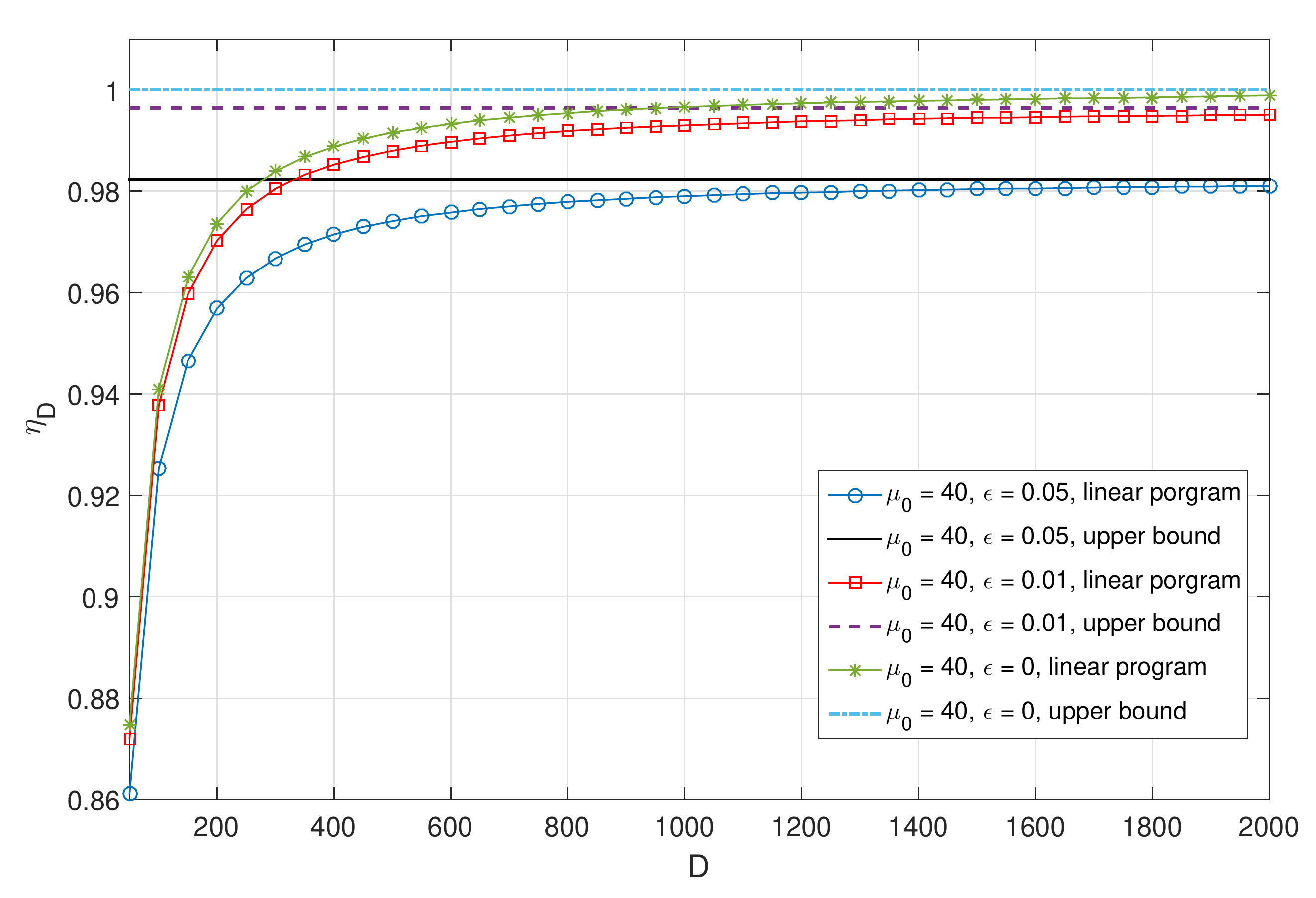}
\caption{The maximum rate efficiency of the optimal degree distribution functions versus the maximum degree for different values of $\epsilon$.}
\label{OptEffFig}
\end{figure}

The following lemma provides an upper bound on the maximal rate efficiency of a Raptor code in the low SNR regime with parameter $\epsilon$.
\begin{lemma}
\label{maxEffProp}
The maximum achievable rate efficiency $\eta$, is upper bounded by:
\begin{align}
\eta<\frac{4\ln(2)}{4\ln(2)+\epsilon}.
\label{maxeffciciency}
\end{align}
\end{lemma}
The proof of this lemma is provided in Appendix \ref{ProofmaxEffProp}. For $\epsilon=0.05$, the maximum efficiency will be 0.9823, which can be achieved when $D$ is very large, tending to infinity. In other words, for a nonzero $\epsilon$, there is always a gap to the capacity, which can be characterized by (\ref{maxeffciciency}) in terms of rate efficiency, $\eta$. Fig. \ref{OptEffFig} shows the maximum rate efficiency of a Raptor code versus the maximum degree. As can be seen in this figure, with increasing the maximum degree, the maximum efficiency get closer to the upper bound provided in (\ref{maxeffciciency}).

\subsection{A Discussion on Determining $\epsilon$}
\begin{figure}[t]
\centering
\includegraphics[scale=0.3]{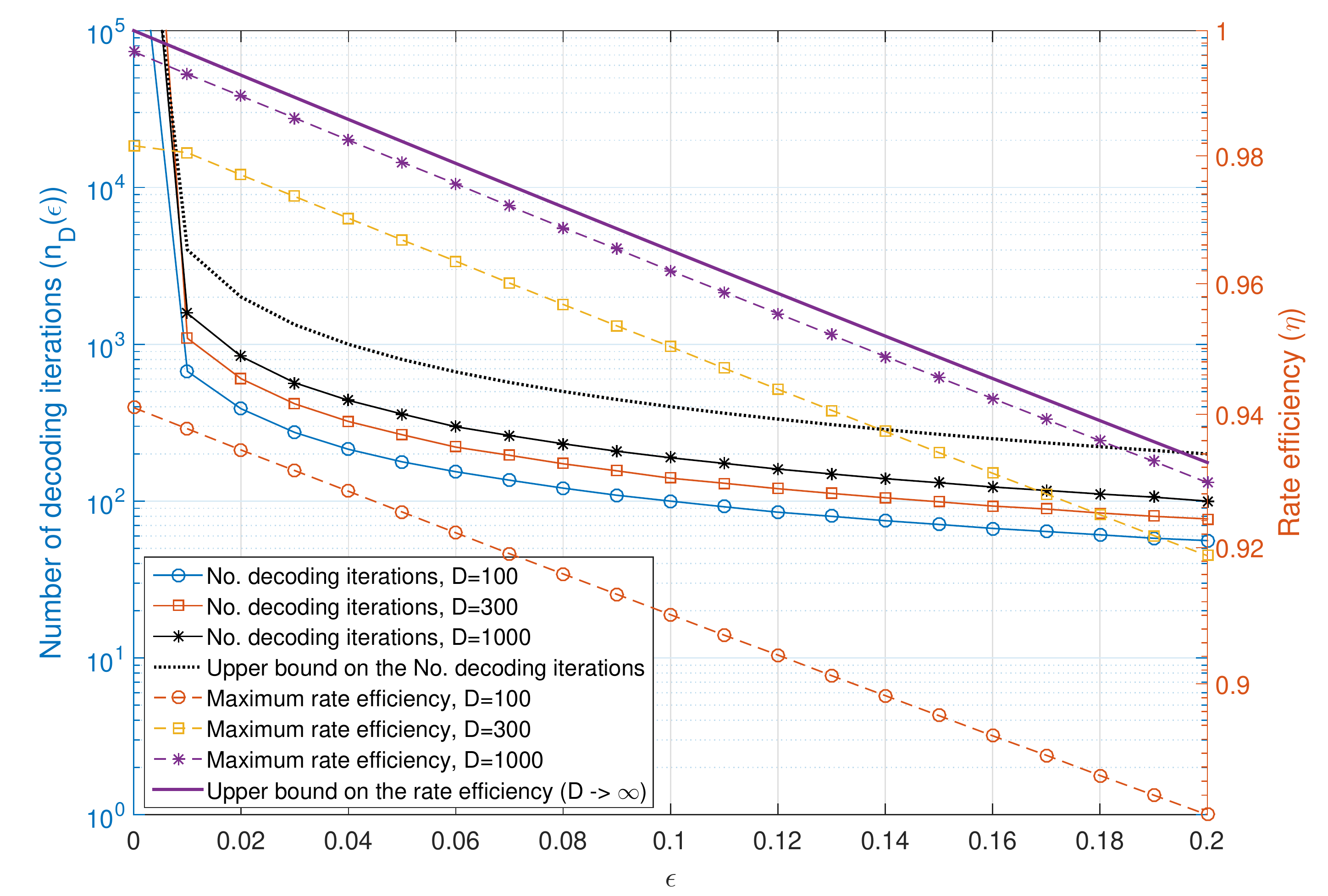}
\caption{The average number of decoding iterations and rate efficiency vs $\epsilon$ for  a Raptor code with optimal degree distributions and different maximum degrees.}
\label{TardeFig}
\end{figure}

The parameter $\epsilon$ in (\ref{optlowsnr}) determines the average number of decoding iterations in the BP decoding algorithm. The maximum number of decoding iterations for a Raptor code with the maximum degree $D$, denoted by $n_{D}(\epsilon)$, can be upper bounded by:
\begin{align}
n_D(\epsilon)\le \frac{\mu_0}{\epsilon},
\label{NoITUp}
\end{align}
which arises from condition ($i$) in (\ref{optlowsnr}), where the minimum gap between the left hand side and the right hand side of the inequality is considered to be $\epsilon$. As can be seen in (\ref{NoITUp}), the maximum number of decoding iterations is proportional to the inverse of $\epsilon$, which can also be seen in Fig. \ref{TardeFig}.
This shows that for practical implementation of Raptor codes in the low SNR regime, a small $\epsilon$ leads to a huge decoding complexity. Moreover, $\epsilon$ can be chosen such that the rate efficiency is high enough with a reasonable decoding complexity. For example as shown in Fig. \ref{TardeFig}, when the number of decoding iterations is set to $1000$, $\epsilon$ must be set to greater than  $0.01$. Although, $\epsilon=0.01$ gives a higher rate efficiency in theory, as will be shown in the next section, it leads to a worse efficiency in actual simulations when the SNR is not zero.

\section{Simulation Results and Discussion}
\begin{figure}[t]
\centering
\includegraphics[scale=0.33]{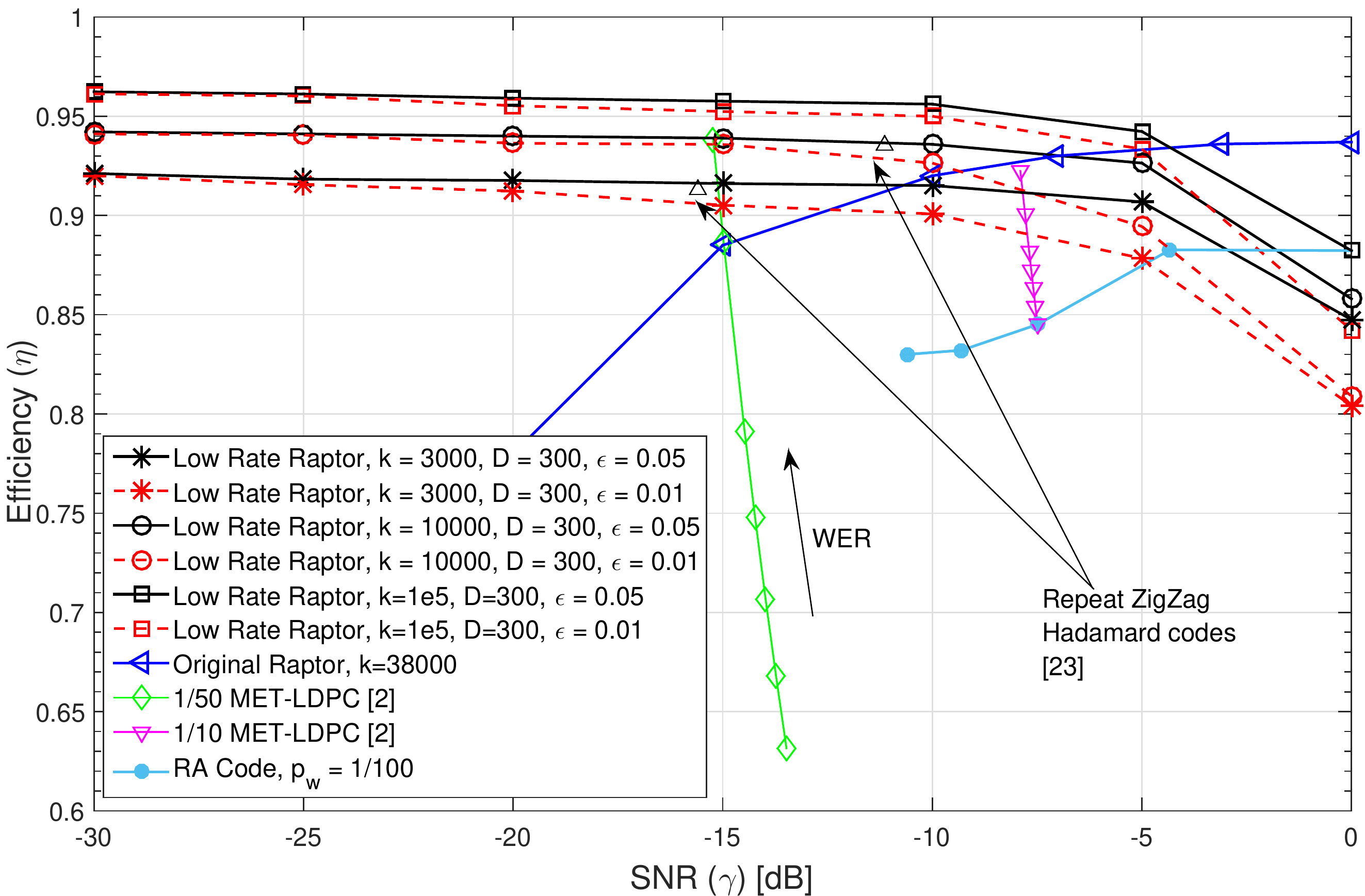}
\caption{Rate efficiency versus the SNR for a Raptor code with the optimized degree distribution and different message lengths,  different values of $D$ and $\epsilon$, when $\mu_o=40$. The code length for RA codes \cite{Sarah_J_RA_QKD} is 64800 and the code length for MET-LDPC is $10^{6}$ \cite{PhysRevA}. The WER for rate 1/50 and 1/10 MET-LDPC codes are respectively $\{0.54,0.16,0.038,0.021,0.014,0.012,0.009\}$ and $\{1.8\times10^{-3},5.7\times10^{-4},1.2\times10^{-4},3\times10^{-5},1.5\times10^{-5}\}$ from top to the bottom.}
\label{EffSingleFig}
\end{figure}

In this section, we carry out simulations on Raptor codes in the low SNR regime. Fig. \ref{EffSingleFig} shows the rate efficiency, $\eta$, of a Raptor code with the optimized degree distribution function versus the SNR when the maximum number of iterations is 1000. As can be seen in this figure, the code achieves higher efficiencies in lower SNRs, which is due to the fact that the degree distributions were designed for the low SNR regime. Moreover, by increasing the message length, the rate efficiency improves which is due to the fact that the tree assumption and independence of the messages in the SPA decoding are more likely to be met as the message length increases. It is important to note that the efficiency at SNRs larger than -10 dB using the degree distributions obtained from Table \ref{OptDegTab} is not high enough which is due to the fact that these degree distributions were designed for very low SNRs. For SNR larger than -10 dB, one could solve the original linear program (\ref{optorig}) to find the optimal degree distribution, which has been shown that can achieve efficiencies larger than 92\% at those SNRs \cite{Raptor_WideSNR}. The rate efficiencies for these codes have been also shown in Fig. \ref{EffSingleFig}

It is also important to note that the choice of $\epsilon$ plays and important role in the efficiency of the Raptor code. More specifically, for finite length code, as can be seen in Fig. \ref{EffSingleFig}, when $\epsilon=0.05$, the code achieves higher efficiencies compared to the code with $\epsilon=0.01$. Although, for infinite length code higher efficiencies can be achieved for a lower $\epsilon$ (Fig. \ref{TardeFig}), in real applications where the message length is finite, $\epsilon$ cannot be very small, since otherwise the decoder does not converge due to the random nature of noise and the code graph.

We have also compared rate 1/10 and 1/50 MET-LDPC codes designed in \cite{PhysRevA} and several RA codes in \cite{Sarah_J_RA_QKD} in Fig. \ref{EffSingleFig}. As can be seen in this figure, Raptor codes with an appropriately designed degree distribution can achieve higher efficiencies in comparison with the fixed rate codes in the entire SNR range. It is important to note that LDPC and MET-LDPC codes can be designed for different rates, but, existing techniques fail to design very high efficiency low rate codes for the low SNR regime. Moreover, the rate efficiency of fixed rate codes decreases with decreasing the word error rate, which is also clear from Fig. \ref{EffSingleFig}. For the sake of comparison, the rate efficiency of a few existing repeat ZigZag Hadamard codes \cite{RepeatZigZagHadamard} have been also plotted in Fig. \ref{EffSingleFig}.

\begin{figure}[t]
\centering
\includegraphics[scale=0.33]{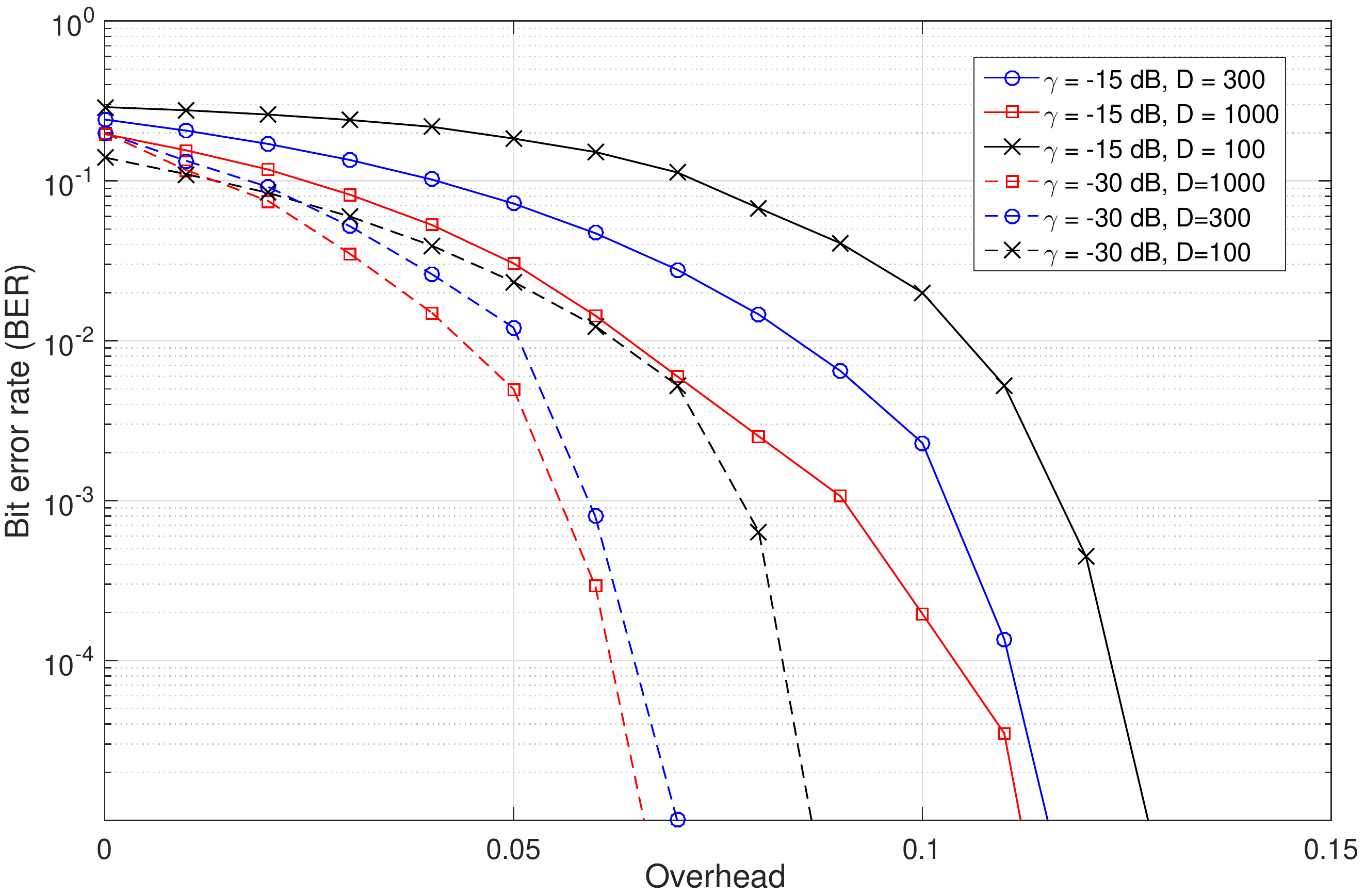}
\caption{Bit error rate versus the overhead for a Raptor code with different degree distributions at different SNRs, when $\mu_o=40$.}
\label{berfig}
\end{figure}

Fig. \ref{berfig} shows the bit error rate versus the overhead for a Raptor code at different SNRs. More specifically, let $k$ and $n$ denote the message length and the number of coded symbols, respectively, then the overhead is defined as $(n-k)/n$. As can be seen in this figure, with increasing the maximum degree, a lower overhead is required to achieve lower BERs. Moreover, at lower SNRs, the overhead required to achieve lower BERs decreases due to the fact that the optimized degree distribution was designed for the low SNR regime, and with decreasing the SNR, the approximations in the linear program (\ref{optlowsnr}) become more realistic.

\section{Conclusions}
This paper provided a detailed study on Raptor codes in the low SNR regimes, where the degree distribution design was formulated and an exact expression for degree distribution function in the asymptotic case, when the maximum degree goes to infinity, was provided. Accordingly, the fraction of output nodes of small degrees were found, which are in a close agreement with the results obtained from the linear programming solution of the optimal degree distribution function. A more practical degree distribution design was also proposed, where the code efficiency and decoding complexity were taken into account when optimizing the degree distribution. An upper bound on the maximum rate efficiency of the designed Raptor code was also proposed. Simulation results show that Raptor code with the designed degree distribution functions closely approach the capacity of the AWGN channel in very low SNRs.
\appendices
\section{Proof of Lemma \ref{feasibility}}
\label{feasibilityproof}
The proof of this lemma follows directly from the proof of Lemma 2 and Theorem 1 in \cite{Raptor_BIAWGN}. More specifically, $f_d(\mu)$ is shown to be increasing with $\mu$ for $d>1$, and it also decreases with $d$ for any $\mu$ and SNR. Therefore, we have:
\begin{align}
\nonumber f_d(\mu)\le f_1(\mu)=2\mathrm{SNR}.
\end{align}
Therefore, for constraint $(i)$ in (\ref{optorig}), we have:
\begin{align}
\nonumber \alpha\sum_{d=1}^{D}\omega_df_d(\mu_i)\le \alpha\sum_{d=1}^{D}\omega_d f_{1}(\mu_i)=\alpha f_{1}(\mu_i)=2\alpha\mathrm{SNR},
\end{align}
thus, we have $\mu_i<2\alpha\mathrm{SNR}$. For the maximum value of $\mu_i$, we also have $\mu_o\le 2\alpha\mathrm{SNR}$. Now suppose that $\mathrm{SNR}>\mu_o/2\alpha$, then it is clear that $\omega(x)=1$ satisfies all constraints in (\ref{optorig}), thus the optimization is feasible.  In an asymptotic case, when the maximum degree is very large, $\alpha$ is potentially infinite, and the optimization problem is feasible for $\mathrm{SNR}>\lim_{\alpha\to \infty}\mu_o/2\alpha = 0$. This completes the proof.
\section{Proof of Proposition \ref{prop1}}
\label{prop1proof}
Statements (a)-(e) were previously proven in \cite{Raptor_BIAWGN,SPA_LDPC_IT}, therefore we only provide the proof for (f). It is clear that for the inverse function we can write:
\begin{align}
\label{invintfo}
\int_{0}^{1-\delta}\varphi^{-1}(t)dt=\varphi^{-1}(1-\delta)-\int_{0}^{\varphi^{-1}(1-\delta)}\varphi(t)dt.
\end{align}
As shown in \cite{SPA_LDPC_IT}, $\varphi(x)$ can be represented by:
\begin{align}
\varphi(x)=1-4\sum_{k=0}^{\infty}(-1)^k \mathrm{e}^{x(k^2+k)}Q\left(\sqrt{\frac{x}{2}}(1+2k)\right),
\label{exactphi}
\end{align}
where $Q(x)=\frac{1}{\sqrt{2\pi}}\int_{x}^{\infty}\mathrm{e}^{-t^2/2}dt$, which can also be calculated using the following integral \cite{QfuncInt}:
\begin{align}
Q(x)=\frac{1}{\pi}\int_{0}^{\pi/2}\mathrm{e}^{-\frac{x^2}{2\sin^2 \theta}}d\theta.
\label{Qfuncint}
\end{align}
By substituting (\ref{exactphi}) and (\ref{Qfuncint}) in (\ref{invintfo}), and since $\varphi^{-1}(1)=\infty$, we have
\begin{align}
\nonumber &\lim_{\delta\to0}\int_{0}^{1-\delta}\varphi^{-1}(t)dt\\
\nonumber&=4\sum_{k=0}^{\infty}\frac{(-1)^k}{\pi}\int_{0}^{\pi/2}\int_{0}^{\infty}\mathrm{e}^{x\left(k^2+k-\frac{(2k+1)^2}{4\sin^2\theta}\right)}dxd\theta\\
\nonumber&=4\sum_{k=0}^{\infty}\frac{(-1)^k}{\pi}\int_{0}^{\pi/2}\frac{\mathrm{e}^{\left.x\left(k^2+k-\frac{(2k+1)^2}{4\sin^2\theta}\right)\right|_{x=0}^{\infty}}}{k^2+k-\frac{(2k+1)^2}{4\sin^2\theta}}~d\theta\\
\nonumber &\overset{(i)}{=}4\sum_{k=0}^{\infty}\frac{(-1)^k}{\pi}\int_{0}^{\pi/2}\frac{4\sin^2\theta}{4(k^2+k)\cos^2\theta+1}~d\theta\\
\nonumber &\overset{(ii)}{=}4\sum_{k=0}^{\infty}\frac{(-1)^k}{\pi}\int_{0}^{\infty}\frac{4u^2}{(2k+1)^2+u^2}\frac{du}{1+u^2}\\
\nonumber &=4\sum_{k=0}^{\infty}\frac{(-1)^k}{\pi}\int_{0}^{\infty}\left(\frac{\frac{(2k+1)^2}{k^2+k}}{(2k+1)^2+u^2}+\frac{\frac{-1}{k^2+k}}{1+u^2}\right)du\\
\nonumber &=4\sum_{k=0}^{\infty}\frac{(-1)^k}{\pi}\left[\frac{2k+1}{k^2+k}\mathrm{atan}\left(\frac{k^2+k}{(2k+1)^2}u\right)-\frac{\mathrm{atan}(u)}{k^2+k}\right]_{u=0}^{\infty}\\
\nonumber &\overset{(iii)}{=}4\sum_{k=0}^{\infty}\frac{(-1)^k}{k+1}\overset{(iv)}{=}4\ln(2),
\end{align}
where step $(i)$ follows from the fact that $k^2+k-\frac{(2k+1)^2}{4\sin^2\theta}<0$, in step $(ii)$ we performed variable change $u=\tan \theta$, step $(iii)$ follows from $\mathrm{atan}(\infty)=\pi/2$, and finally step $(iv)$ follows from the Maclaurin series of $\ln(1+x)$ for $x=2$. This completes the proof.

\section{Proof of Lemma \ref{generallemma}}
\label{generallemmaproof}
For the first condition, let us assume that there is a degree distribution function of maximum degree $D$, which satisfies the following inequality for $0\le x \le1$, i.e., $\delta=0$:
\begin{align}
\Omega'(x)\ge \frac{1}{4\ln(2)}\varphi^{-1}(x),
\end{align}
then it is clear that $\Omega'(x)$ must goes to infinity when $x\to 1$, as $\lim_{x\to 1}\varphi^{-1}(x)=\infty$. This contradicts the fact that the maximum degree $D$ is finite. This proves the first condition. The second condition is directly proven from the fact that by increasing the maximum degree the optimization problem will have a larger degree of freedom.
\section{Proof of Theorem \ref{theorem1}}
\label{theorem1proof}
For the optimal degree distribution $\Omega^{(D)}(x)$ with maximum degree $D$, we have:
\begin{align}
\label{firstcond}
 \Omega^{(D)'}(x)\ge\frac{1}{4\ln(2)}\varphi^{-1}(x)dt, ~~0\le x\le1-\delta_D,
\end{align}
and accordingly
\begin{align}
\label{secondcond}
\Omega^{(D)}(x)\ge\frac{1}{4\ln(2)}\int_{0}^{x}\varphi^{-1}(t)dt, ~~0\le x\le1-\delta_D.
\end{align}
We then define function $\Delta^{(D)}(x)$ as follows:
\begin{align}
\Delta^{(D)}(x)=\Omega^{(D)}(x)-\frac{1}{4\ln(2)}\int_{0}^{x}\varphi^{-1}(t)dt,
\end{align}
which is obviously an increasing function of $x$ due to (\ref{firstcond}) for $0\le x\le1-\delta_D$. Therefore, we have:
\begin{align}
0\le\Delta^{(D)}(x)\le \Delta^{(D)}(1-\delta), ~~0\le x\le1-\delta_D.
\end{align}
According to Lemma \ref{generallemma}, $\delta_D$ is a decreasing function of $D$, therefore, we have:
\begin{align}
\nonumber &\lim_{D\to \infty}\Delta^{(D)}(1-\delta_D)\\
\nonumber &=\lim_{D\to \infty}\Omega^{(D)}(1-\delta_D)-\lim_{D\to \infty}\frac{1}{4\ln(2)}\int_{0}^{1-\delta_D}\varphi^{-1}(t)dt=0\\
\nonumber &\le1-\frac{1}{4\ln(2)}\int_{0}^{1-\delta_{\infty}}\varphi^{-1}(t)dt,
\end{align}
which arises from the fact that $\Omega(x)\le1$ for $x\in[0,1]$. This completes the proof.
\section{An Alternative Approach to Find the Optimal Degree Distribution in the Asymptotic Case}
\label{altproof}
By using  Lemma \ref{generallemma}, it can be verified that $g(x)\triangleq\frac{1}{4\ln(2)}\varphi^{-1}(x)$ is a probability distribution function defined in $[0,1)$, as it is always positive and its integral over $[0,1]$ is 1. Therefore, we can find the $n^{th}$ moments of $g(x)$, referred to as $\mathcal{G}_n$, as follows:
\begin{align}
\mathcal{G}_n=\int_{0}^{1}t^ng(t)dt, ~~n=0,1,\ldots.
\label{alt1}
\end{align}
Also, from Theorem \ref{theorem1} and assuming that the optimization problem (\ref{optorig}) is always feasible for $D$ being sufficiently large and when $\gamma$ goes to zero, we have:
\begin{align}
\nonumber \Omega^{(\infty)}(x)=\frac{1}{4\ln(2)}\int_{0}^{x}\varphi^{-1}(t)~dt, ~0\le x \le 1,
\end{align}
and equivalently,
\begin{align}
\Omega^{(\infty)'}(x)=g(x), ~0\le x < 1.
\label{alt3}
\end{align}
By substituting (\ref{alt3}) into (\ref{alt1}), we have:
\begin{align}
\mathcal{G}_n=\int_{0}^{1}t^n\sum_{d=1}^{\infty}d\Omega_d ~t^{d-1}dt=\sum_{d=0}^{\infty}\frac{d}{d+n}\Omega_d,
\end{align}
which can be also represented as follows:
\begin{align}
\frac{1-\mathcal{G}_n}{n}=\sum_{d=0}^{\infty}\frac{1}{d+n}\Omega_d.
\end{align}
The above equation can be represented in matrix form as follows:
\[\left[\begin{array}{c}
\frac{1-\mathcal{G}_1}{1}\\
\frac{1-\mathcal{G}_2}{2}\\
\frac{1-\mathcal{G}_3}{3}\\
\vdots\end{array}\right] = \left[\begin{array}{c c c c}
1& \frac{1}{2}& \frac{1}{3}& \cdots\\
\frac{1}{2}& \frac{1}{3}& \frac{1}{4}& \cdots\\
\frac{1}{3}& \frac{1}{4}& \frac{1}{5}& \cdots\\
\cdots&\cdots&\cdots&\ddots\end{array}\right]\left[\begin{array}{c}
\Omega_0\\
\Omega_1\\
\Omega_2\\
\vdots\end{array}\right]\]
where the square matrix in the above equation is the Hilbet matrix, where the $i^{th}$ entry in the $j^{th}$ column, denoted by $h_{i,j}$, is $h_{i,j}=1/(i+j-1)$ \cite{Hilbert}. The entries of the inverse Hilbet matrix, denoted by $H^{-1}$, can be calculated as follows:
\begin{align}
\nonumber&(H^{-1})_{i,j}=\\
\nonumber&(-1)^{i+j}(i+j-1)\dbinom{n+i-1}{n-j}\dbinom{n+j-1}{n-i}\dbinom{i+j-2}{i-1}^2,
\end{align}
where $n$ is the order of the Hilbert matrix \cite{Hilbert}. Therefore, the optimal degree distribution can be calculated as follows:
\begin{align}
\Omega_d=\sum_{j=1}^{\infty} (H^{-1})_{d+1,j}\frac{1-\mathcal{G}_j}{j}.
\end{align}
\section{Proof of Lemma \ref{avgdegProp}}
\label{ProofavgdegProp}
Condition (i) in (\ref{optlowsnr}) must hold for $x_0=\varphi^{-1}(\mu_0)$ for a capacity approaching code. As $\Omega(x)$ and therefore $\Omega'(x)$ are continuous and monotonically increasing  with $x$, we have:
\begin{align}
\nonumber\sum_{d=1}^Dd\Omega_d\ge\sum_{d=1}^{D}d\Omega_dx_0^{d-1}\ge \frac{\eta(\varphi^{-1}(x_0)+\epsilon)}{4\ln(2)}
\end{align}
which directly results in (\ref{avgdegree}).

\section{Proof of Lemma \ref{maxEffProp}}
\label{ProofmaxEffProp}
From (\ref{optlowsnr}), we have:
\begin{align}
\nonumber\frac{4\ln(2)}{\eta}\int_{0}^{\varphi(\mu_0)}\sum_{d}d\Omega_{d}x^{d-1}dx>\int_{0}^{\varphi(\mu_0)}(\varphi^{-1}(x)+\epsilon)dx,
\end{align}
which can be further simplified as follows:
\begin{align}
\nonumber\frac{4\ln(2)}{\eta}\sum_{d}\Omega_{d}\left[\varphi(\mu_0)\right]^{d}dx&>\mu_0-\int_{0}^{\mu_0}\varphi(x)dx+\epsilon\varphi(\mu_0).
\end{align}
Since $\lim_{x\to \infty}\varphi(x)=1$ and $\sum_{d=1}^D\Omega_d=1$, we can then rewrite the above inequality as follows:
\begin{align}
\frac{4\ln(2)}{\eta}>\lim_{\mu_0\to \infty}\left(\mu_0-\int_{0}^{\mu_0}\varphi(x)dx\right)+\epsilon.
\label{simmaxeff}
\end{align}
According to lemma \ref{generallemma}, we have:
\begin{align*}
\lim_{\mu_0\to \infty}\left(\mu_0-\int_{0}^{\mu_0}\varphi(x)dx\right)=\int_{0}^{1}\varphi^{-1}(t)dt=4\ln(2),
\end{align*}
therefore, (\ref{simmaxeff}) can be rewritten as follows:
\begin{align*}
\frac{4\ln(2)}{\eta}>4\ln(2)+\epsilon,
\end{align*}
which directly results in (\ref{maxeffciciency}).

\bibliographystyle{IEEEtran}
\footnotesize
\bibliography{IEEEabrv,sample2}

\end{document}